\documentclass[12pt,letterpaper]{iopart}
\usepackage{graphicx}
\usepackage{bm}
\begin{document}
\title{An investigation into the feasibility of myoglobin-based single-electron transistors}
\author{Debin Li$^1$, Peter M. Gannett$^2$, and David Lederman$^1$}
\address{$^1$ Department of Physics, West Virginia University, Morgantown, WV, 26506-6315, USA}
\address{$^2$ Department of Basic Pharmaceutical Sciences, West Virginia University, Morgantown, WV, 26506-9500, USA}
\ead{david.lederman@mail.wvu.edu}

\begin{abstract} 
Myoglobin single-electron transistors were investigated using nanometer-gap platinum 
electrodes fabricated by electromigration at cryogenic temperatures.  Apomyoglobin 
(myoglobin without heme group) was used as a reference.  The results suggest single 
electron transport is mediated by resonant tunneling with the electronic and vibrational levels of the 
heme group in a single protein.  They also represent a proof-of-principle that proteins with redox centers 
across nanometer-gap electrodes can be utilized to fabricate single-electron transistors.  
The protein orientation and conformation may significantly affect the conductance of these devices.
Future improvements in device reproducibility and yield will require control of these factors.  
\end{abstract}
\pacs{ 73.23.Hk, 73.23.-b, 85.35.Gv, 82.37.Rs, 87.14.E-, 87.80.Nj}
\submitto{Nanotechnology}
\maketitle
\section{Introduction}
Electron transfer is of fundamental importance to proteins whose functionality relies on 
redox reactions with other proteins, cofactors, or the external environment.  In proteins 
containing heme groups, a single Fe ion in a protoporphyrin ring changes valence state 
between ferrous (Fe$^{2+}$) and ferric (Fe$^{3+}$) oxidation states as an electron is accepted by 
or donated to the protein~\cite{meunier:2004}.  Much experimental work has been performed to develop 
an understanding of protein electron transfer mechanisms \cite{marcus:1985,bendall:1996} most of it based on 
statistical averages of protein ensembles measured by spectroscopic and 
electrochemical techniques \cite{tao:1996,chi:2005}.  

Electron transfer through individual molecules has been probed recently using 
electrochemical scanning tunneling microscopy (EC-STM) \cite{tao:1996,alessandrini:2006}.  In EC-STM, a gate 
voltage is used to align a molecule’s redox level to the Fermi levels of the tip and 
substrate with the sample in solution at room temperature.  The results are explained 
using resonant tunneling \cite{tao:1996,alessandrini:2006} and two-step electron transfer models \cite{chi:2005}.  
According to resonant tunneling models, electrons 
transfer via resonant tunneling due to alignment of molecular orbital energy levels with 
the Fermi level of the contact electrodes.  In the two-step process, conformational 
changes in the molecule occur during tunneling which change the orbital energy levels.  
This process is thought to be especially prevalent in redox molecules \cite{alessandrini:2006,alessandrini:2005}.  
In a related group of experiments, measurements on large assemblies of localized proteins using mercury contacts have 
been used to distinguish between tunneling and hopping electron transfer mechanisms \cite{ron:2010}.

Nanometer-gap electrodes fabricated by electromigration techniques \cite{park:1999} have made it 
possible to study electron conductance through small single redox molecules.  By 
applying a bias voltage $V_B$ between electrodes and a gate voltage $V_G$ to the sample's substrate, 
molecular energy levels can be probed and characteristic single-electron transistor 
(SET) behavior, such as a Coulomb blockade or a Kondo resonance, can be 
observed~\cite{park:2002,kubatkin:2003,yu:2004,vanderzant:2006a}.  In particular, these
experiments have demonstrated that molecular junctions fabricated in this way can be used
to probe the electron transfer properties of the molecules.  In these measurements, 
the presence of Coulomb blockade phenomena by itself is not proof that
the molecular properties are being probed because the characteristics of the junction itself after
the breaking process can affect the measurements~\cite{vanderzant:2006}, and therefore
control experiments must be performed to validate the data.  These measurements, if performed correctly,
rely on a resonant tunneling process which in principle indicate that tunneling is a possible
electron transfer mechanism.  Our aim in this work is to determine whether
tunneling is a viable electron transfer mechanism in redox proteins such as myoglobin
by attempting to fabricate and measure single electron transistors.   

Here we describe the study of single-electron transistors composed of the protein 
myoglobin (Mb).  Mb is a roughly spherical globular metalloprotein containing a heme 
group surrounded by a folded single-chain polypeptide consisting of 153 amino acids 
(apomyoglobin) \cite{garrett:1999}.  Mb was chosen because 1) it is relatively small ($4.4\textrm{ nm }\times 4.4\textrm{ nm }\times 2.5\textrm{ nm}$), 
which is desirable to increase electron tunneling probability, 2) the apo 
form is readily available and well studied and can thus be used as a control to discern 
effects related to the heme group, and 3) its structure and vibrational spectra are well 
known.  In other words, Mb is a model system with which to test bioelectronic 
devices based on heme-containing proteins.  

This paper is organized as follows.  After briefly discussing the general properties of 
myoglobin and single-electron transistors in the \textit{Background} section, 
we describe the experimental techniques used,
including break junction fabrication, electrical measurements, and protein incorporation.
In the \textit{Results} section we describe results from bare junctions, 
junctions with apomyoglobin (apoMb),
and junctions with Mb.  In the \textit{Discussion} section we analyze our
data using the known properties of myoglobin and the results from the bare and 
apoMb samples. Because
SET conductance can result from metal grains produced by the electromigration 
process~\cite{houck:2005,vanderzant:2006}, an important distinction between
samples with and without protein incorporation is the observation of
inelastic tunneling from discrete vibrational modes~\cite{park:1999,kubatkin:2003,chae:2006}.
We also identify heme-group resonant tunneling by comparing the behaviors of apoMb samples with Mb samples.  
The \textit{Conclusions} section summarizes our results and identifies areas where the experiments can
be improved.  In \textit{Appendix A} we describe differential SET conductance calculations with and
without tunneling-induced structural changes in the protein.

\section{Experimental techniques}
\subsection{Break-junction fabrication}
\begin{figure}
\includegraphics{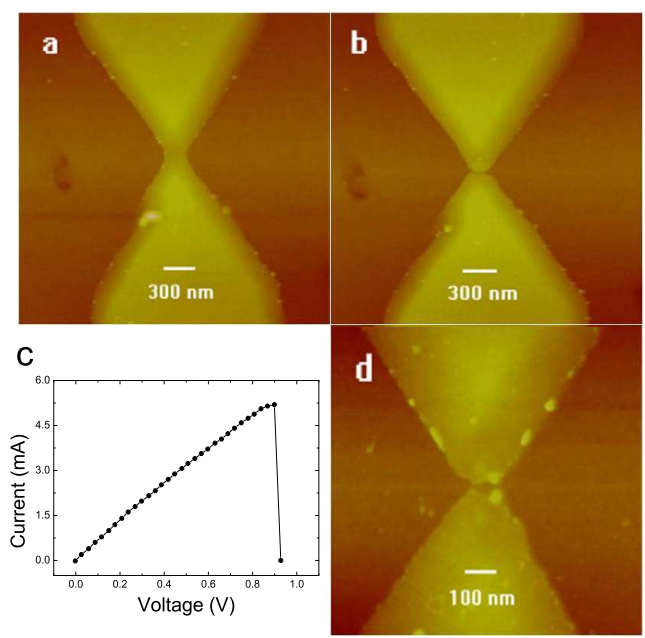}
\centering
\caption{\label{fig:junctions}(a) Atomic force microscopy (AFM) image for one Pt junction which 20 nm thickness in large area, 8 to 10 nm in the middle. 
(b) AFM image after breaking the junction by electromigration at room temperature. 
(c) $I-V$ measurement for breaking the junction at 0.9 V and room temperature.  
(d) AFM image for another junction broken by an electrical pulse (-50 mV) at $T = 6$~K. }
\end{figure}

The junctions were grown on Si substrates with a SiO$_2$ layer 200 nm thick.   
The resistivity of the degenerately-doped silicon wafers 
ranged from 0.002~$\Omega$-cm to 0.003~$\Omega$-cm.  The Si substrate served as the gate 
contact. The widths of junctions were usually between 100~nm and 300~nm.  
The junctions were fabricated by electron-beam lithography using a lift-off technique.  Pt  films, 20 nm thick, were 
deposited on a Si/SiO$_2$ substrate by magnetron sputtering with approximately 
a $15^\circ$ deposition angle with respect to the surface normal of the sample.  
The samples were rotated about the surface normal during deposition.  
The Pt deposition rate was 0.028 nm/s and the base pressure was 
$5\times 10^{-8}$~Torr; an Ar pressure of 1.5 mTorr was used during deposition.  
The  junction thickness at its narrowest point 
was approximately 8~nm to 10~nm because of a shadow effect 
during the sputtering deposition process, as shown in figure~\ref{fig:junctions}(a).  
The junction with this kind bow-tie shape tends to break only in the middle, 
does not require high voltages to break [figures~\ref{fig:junctions}(b) and (c)], 
and can be broken by an electrical pulse with a low negative voltage, as shown in 
figure \ref{fig:junctions}(d)~\cite{hayashi:2008}.  
The total resistance of the junctions (including parasitic resistance) at room temperature 
ranged between 100~$\Omega$ and 2000~$\Omega$.  The narrowest gap between electrodes formed in this way
are on the order of 5~nm, as shown in figure~\ref{fig:junctions}.

\subsection{Protein incorporation and viability}
Mb and apoMb at least 95\% pure were obtained from horse skeletal muscle (Sigma) in its powder 
state and essentially salt-free.  The protein powders were dissolved in a 10 mM phosphate 
buffer saline (PBS, Sigma, $\textrm{pH} = 
7.4$) with 14.2 mM and 31.5 $\mu$M concentrations for Mb and apoMb, respectively.  
Protein solutions were passed through a 
0.2~$\mu$m pore size filter to remove larger protein clusters that may have formed. 
In order to remove possible contaminants from 
the electrode surface, junctions were rinsed with optima acetone and isopropanol and 
subsequently cleaned with O$_2$ plasma.  After the cleaning procedure, 
the protein solution was dropped on the samples.  
The samples were kept at 4~$^\circ$C in order to let water evaporate.  A drop of the solution was placed on the 
unbroken junction prior to cooling down the samples.  

Linker molecules were avoided because junctions exposed to them 
(without protein) generated resonant tunneling electronic data. 
These data showed a weak dependence on gate voltage, which would have made it difficult
to disntiguish the protein behavior from that of the linker (see supplementary information). 
Raman spectra showed that the Mb was active after 
cooling down to $T=5$~K and then warming up to room temperature~\cite{li:2009}.  
Therefore, it is unlikely that the protein denatured 
during the cooling process.  Another possible problem is that the local temperature during the electromigration process 
might be high enough to denature the protein \cite{esen:2005,taychatanapat:2007,ward:2008}.  
Measurements of the electrical resistance at low voltages as a 
function of temperature yielded an estimate of a maximum temperature of approximately 400~K, which is
consistent with estimates of 495 K for junctions broken at room temperature~\cite{dong:2006}.    Additional 
Raman scattering measurements performed on protein heated up in air in a furnace or hot plate demonstrated
that myoglobin is viable after heating to 400~K (see supplementary information).  

\begin{figure}
\centering
\includegraphics{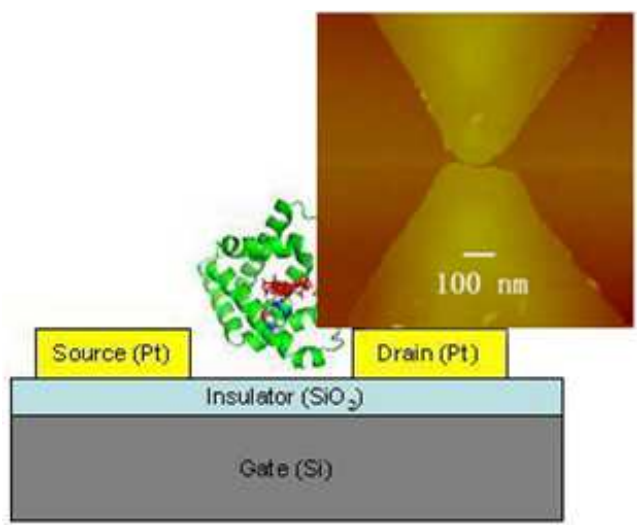}
\caption{\label{fig:Device} Sketch of three-terminal device and AFM image of a bare Pt junction broken by electromigration.  
The AFM image indicates that the gap is on the order of 5 nm wide.}
\end{figure}

The junctions with a bow-tie configuration were fabricated from Pt thin films, 8 to 10 nm thick in the middle, 200 nm 
wide and deposited on a Si wafer with a 200 nm thick SiO$_2$ insulating layer.  The doped Si substrate was used
as a gate contact in all samples, as shown in figure~\ref{fig:Device}.  Electrical measurements were performed 
immediately after breaking the junction at low temperature (either 77 K or 6 K).  
The conductance ($dI/dV$) as a function of 
$V_B$ and $V_G$ was acquired by numerically differentiating current-bias voltage ($I\textrm{-}V$) curves.  
Junctions were measured without any protein, with apomyoglobin, and with myoglobin.  

\section{Results}

\subsection{Bare junctions}
\begin{figure}
\centering
\includegraphics{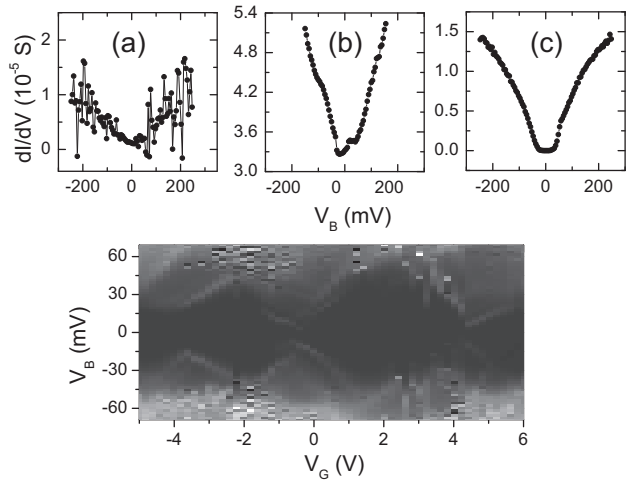}
\caption{\label{fig:blanks} Typical results for bare Pt junctions.  (a) Conductance as a function of bias voltage measured at 
$T=77$~K.  No conductance peaks are found near zero bias.  (b) Conductance of the 
only sample out of 107 junctions broken at $T=77$~K that had noticeable conductance 
features.  (c) Typical differential conductance at $T=5.5$~K for bare junctions. No conductance 
peaks are evident.  (d) SET behavior measured in a bare Pt junction at $T=5$~K 
[gray scale from $-3.5 \times 10^{-7}$~S (black) to $1.8\times 10^{-5}$~S 
(white)].}
\end{figure}

In the control experiments with bare junctions (without protein), 107 Pt junctions were broken by electromigration at 
$T= 77$~K.  The broken junctions had a resistance ranging from 1~M$\Omega$ to 
1~G$\Omega$.  All but one of the broken junctions did not show differential 
conductance peaks.  Figure~\ref{fig:blanks}(a) is a typical result 
measured at $T = 77$~K.  The conductivity was relatively low and there was a 
significant amount of noise for these structures.  Only one sample had conductance 
peaks, as shown in figure~\ref{fig:blanks}(b).  In addition, 31 bare Pt junctions were 
broken by electromigration and measured at $T=5.5$~K.  Seventeen of these did not 
have conductance peaks as shown in figure~\ref{fig:blanks}(c).  Out of the remaining  
fourteen junctions that had conductance peaks, eleven broken junctions did not have 
a dependence on the gate voltage.  The remaining three junctions did have 
conductance peaks whose positions were dependent on the gate voltage.  Coulomb 
diamonds, characteristic of SET behavior, were observed in the conductance as a 
function of gate and bias voltages as shown in figure~\ref{fig:blanks}(d).  This 
demonstrates that occasionally Pt grains are formed, as has been previously observed 
in Au junctions, which can mask the molecular 
information~\cite{houck:2005,vanderzant:2006,ward:2008}.  These islands have several energy 
levels and the conductance peaks from metal grains can be tuned by gate voltage 
modulation.  Many Coulomb diamonds were observed, each corresponding to the 
addition of an electron to the island.  It is important to note that the appearance
of several Coulomb diamonds is unlikely to happen in single-protein junctions, since 
excessive charging can denature the protein.  

\subsection {Apomyoglobin experiments}

\begin{figure}
\centering
\includegraphics{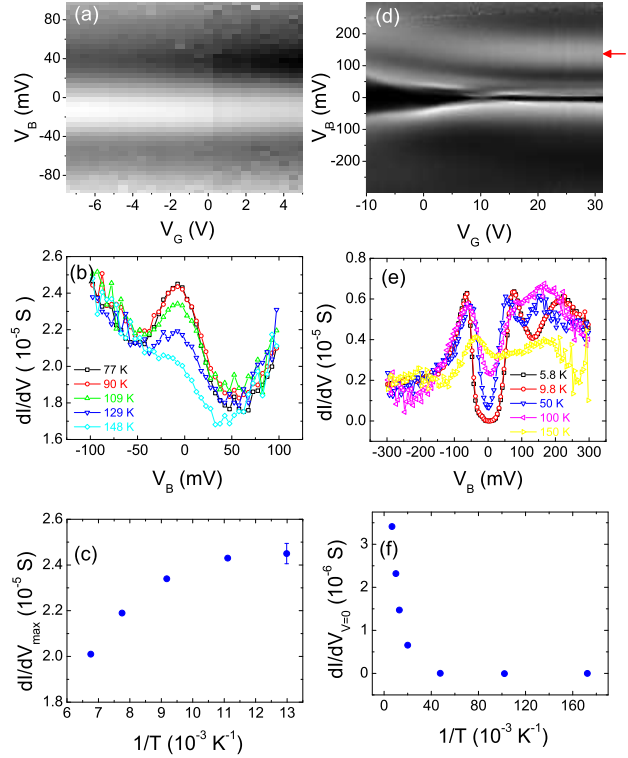}
\caption{\label{fig:apomyodata} ApoMb results. (a) Differential conductance ($dI/dV$) measured at $T=77$~K as a function of bias and 
gate voltages.  Gray scale from $1.56\times 10^{-5}$~S (black) to  $2.64\times 10^{-5}$~S (white) . (b) $dI/dV$ measured at $V_G=0$ for various temperatures. (c) $dI/dV$ of peak in (b) as a function of $1/T$.
(d) $dI/dV$ for another Mb sample [gray scale from 0 (black) to $5.0\times 10^{-6}$~S (white)] measured at $T= 6$~K.  
The red arrow points to a possible vibration-assisted conduction line.  (e) $dI/dV$ at $V_G=0$ for various 
temperatures.  (f) Conduction at $V_B=0$ and $V_G=0$ for sample in (d) as a function of $1/T$. }
\end{figure}

In the apoMb experiments, out of 125 Pt protein-coated junctions broken  
at $T=77$~K, nine had conductance peaks with a very weak response to $V_G$.  
Figure~\ref{fig:apomyodata}(a) is the conductance measured at $T=77$~K 
for one of these apoMb samples.  One conductance peak was approximately constant as a function of $V_G$.  This
peak, measured at $V_G=0$ as a function of temperature, is shown in 
figure~\ref{fig:apomyodata}{b}.  Figure~\ref{fig:apomyodata}(c) shows how the 
peak height depended on temperature.  
The monotonic decrease of the peak height with increasing temperature, together with 
saturation at low temperatures ($T<100\textrm{ K}$) indicates that tunneling, 
and not Mott hopping, is the dominant electron
conduction mechanism in this device~\cite{fowler:1986,azbel:1984}.  
Tunneling is observed because the conductance is smaller than the conductance 
quantum of $G_\circ = 2e^2 /h = 7.75 \times 10^{-5}$~S.  The lack of 
dependence of $V_G$ means, however, that the separation of energy levels 
inside the protein are much larger than $k_BT$~\cite{kastner:2000}.

Forty-three apoMb samples were measured at $T\approx 6$~K.  Fifteen of them had 
conductance peaks and two had a very 
weak $V_G$ dependence.  For three of the apoMb samples, conductance lines changed slightly with $V_G$ as shown in 
figure~\ref{fig:apomyodata}(d).  The spectra measured at $V_G=0$ at several temperatures is shown in 
figure~\ref{fig:apomyodata}(e).  The zero conductance region near $V_B=0$ may arise 
from a Coulomb blockade effect, but the 
lack of Coulomb triangles indicates that the energy levels in the apoMb protein are too close in energy to be able to 
observe a Coulomb staircase.  The temperature dependence of the the conductance at $V_B=0$ is shown in 
figure~\ref{fig:apomyodata}(f).  For temperatures below $T=20$~K the conductance is independent of temperature, 
characteristic of a tunneling process, but at higher temperatures the conductance quickly increases until the spectrum 
completely loses its tunneling gap above $T=100$~K.
The line that starts at $V_B\approx 132.5$~mV, indicated by the red arrow, 
may arise from the vibrational spectra of amino acid 
redox centers~\cite{shih:2008}.

\subsection{Myoglobin experiments}
A total of 215 Mb samples were measured at $T\approx 5$~K, out of which 21 had conductance 
peaks that strongly depended on $V_G$.  In another approach, junctions 
were broken by a single electrical pulse with a very low negative voltage with a 
mechanism similar to voltage-pulse-induced electromigration~\cite{hayashi:2008}.  These thin ($\sim 8$~nm) Pt junctions can be 
broken by a low negative voltage $\approx -60$~mV which 
avoids potentially heating the sample to high temperatures.  Out of 64 Pt Mb-coated junctions broken by 
electrical pulse at $T=6$~K, four had an obvious dependence on $V_G$.  

Of the 25 samples that had gate voltage-dependent behavior, the conductance maps can be divided into
three types:  1)  data where a Kondo-like resonance is observed, that is, conductance near zero bias voltage
that is independent of gate voltage; 2) Coulomb blockade without a two-step process; and 3) Coulomb blockade
with a two-step process.  

\begin{figure}
\centering
\includegraphics{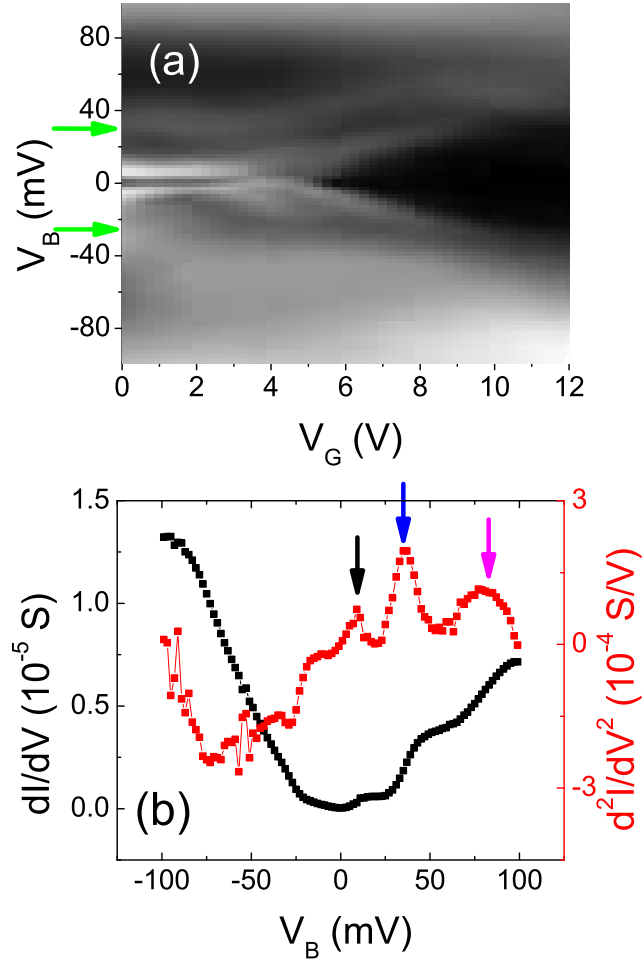}
\caption{\label{fig:myo1data} Mb results. (a)  $dI/dV$ as function of $V_B$ and $V_G$ at $T=6$~K [gray scale is 0 
(black) to $1.33\times 10^{-5}$~S (white)]. Green arrows point to conduction lines 
corresponding to Fe-His vibration-assisted tunneling.  (b) $dI/dV$ (black) and 
$d^2I/dV^2$ (red) spectra at $V_G = 11$~V. Vertical arrows correspond to inelastic tunneling peaks. }
\end{figure}

Figure~\ref{fig:myo1data} shows 
results from one of the samples with a Kondo-like resonance.  When the junction was initially broken by a 
negative electrical pulse, no conductance peaks were found.  However, there 
was a sudden increase in the current when the voltage was increased again, 
indicating a modification of the contact and possible protein incorporation in the gap of electrodes~\cite{park:2002},
although this could also be due to Joule heating resulting from a sudden narrowing of the junction in the bowtie 
region~\cite{strachan:2005}.  Conductance peaks were 
observed thereafter.  In the differential conductance map, shown in figure~\ref{fig:myo1data}(a), 
two intense peaks around $V_B=0$ are independent of $V_G$ for $0 < 
V_G < 3$~V.  For higher $V_G$ values, the peaks are composed of half a Coulomb blockade-like 
triangle.  This could be an indication of a Kondo-resonance assisted tunneling~\cite{park:2002}, although verification 
would require measurement in large magnetic fields.  The 
two lines indicated by the green arrows, corresponding to green arrows in figure~\ref{fig:myo1data}(a), 
do not depend strongly on $V_G$ for $0 < V_G < 1$~V.  The conductance feature starting at 
$V_B\approx 30$~mV could arise from vibrational excitations as this energy coincides with the 
energy of the Fe-His vibration (27.3 meV).  
Conductance steps are also present 
as indicated in figure~\ref{fig:myo1data}(b).  These steps result in peaks in $d^2I/dV^2$, indicated by the 
black (9 mV), blue (35 mV), and magenta ($\approx 80$~mV) arrows in figure~\ref{fig:myo1data}(b).  Taking into 
account that modes increase in energy at low $T$, the arrows correspond to heme-
doming~\cite{rosca:2000,gruia:2008}, Fe-His, and, heme-group deformation (83.7 meV) modes, 
respectively~\cite{feng:2001}.  

\begin{figure}
\centering
\includegraphics{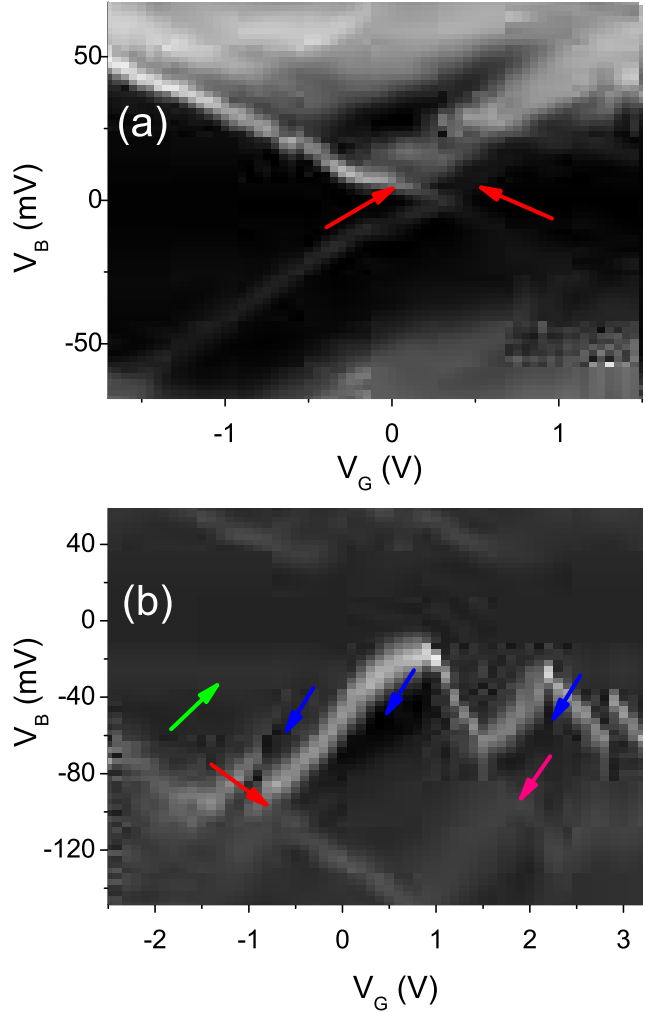}
\caption{\label{fig:myo2data}  (a) Differential conductance ($dI/dV$) data from a Mb sample 
[gray scale from 0 (black) to $5.0\times 10^{-6}$~S (white)] at $T= 6$~K.  
The red arrows point to vibration-assisted conduction lines.  (b)  Differential conductance $T = 6$~K 
[gray scale from $-0.27 \times 10^{-6}$~S (black) to $1.6 \times 10^{-6}$~S (white)]
for another Mb sample.  The blue arrows point to regions of negative differential conductance (NDC).  
The green arrow points to a faint conductance line independent of $V_G$ .  
The red arrow points to another conductance line 
possibly arising from a vibrational-assisted tunneling process or an excited electronic level in the protein(s).}
\end{figure}

Conductance data for a sample which displayed Coulomb blockade behavior
without a two-step process are shown in figure~\ref{fig:myo2data}(a).  
Figure~\ref{fig:Calculations}(b) is a calculation using reasonable 
parameters that reproduce the main characteristics of the experimental data.  The 
appearance of two wide conductance lines starting at $V_B\approx 5.2$~mV, indicated by red 
arrows in figure~\ref{fig:myo2data}(a), arise from inelastic tunneling with the 4.96 meV vibrational mode 
of the heme group~\cite{rosca:2000,gruia:2008}.

Another sample with SET behavior is shown in figure~\ref{fig:myo2data}(b).  
In this case, two deformed Coulomb blockade triangles 
near zero bias voltage can be found without a degeneracy point in the differential conductance graph.  
This means that two charge states do not have the same energy,
and in this case the separation between the top triangles is $>60$ mV.  
This separation, as shown by the calculation in figure~\ref{fig:Calculations}(c), is an indication that the two-step 
electron transfer process is a viable mechanism.  In addition, there are three areas that exhibit 
negative differential conductance (NDC), indicated by the three blue arrows.
One explanation of how NDC behavior may arise is from the existence of two quantum dots in series .  
This suggests that these measurements were the result of transport through two or more 
myoglobin proteins in series with each other.  
The faint line at $V_B \approx -25$~mV, indicated by the green arrow, does not have a strong dependence on $V_G$, 
which indicates electron tunneling by the medium around the heme group.   
We also note that the conductance feature that starts at $V_B = -97$~mV, indicated by the red 
arrow, could arise from one of the excited electronic level of the protein or a vibrational excitation of the protein.  
In particular, this energy coincides with the energy of the tryptophan amino acid in 
Mb which is known to have an energy of $760\textrm{ cm}^{-1} = 94.2$~mV 
(measured at room temperature) \cite{wei:2008}.  The wide conductance 
line that starts at $V_B\approx 93$~mV, indicated by the pink arrow, is another line that 
could arise from the excited state or vibrational excitation.

\section{Discussion}
\begin{figure}
\centering
\includegraphics{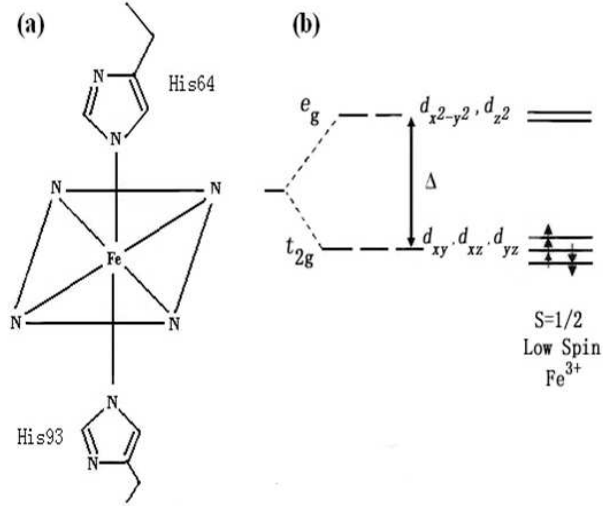}
\caption{\label{fig:Heme} Heme group structure (a) and energies of Fe 
$d$-orbitals in dry metMb (b). }
\end{figure}

Here we discuss our results in terms of the known properties of myoglobin.  
Figure~\ref{fig:Heme}(a) shows the heme group structure of dry Mb.  The fifth Fe ligand binds 
covalently with amino acid His 93, and therefore the Fe-His vibrational mode (27.3 
meV) is an excellent heme-group indicator~\cite{rosca:2000}.  The sixth Fe ligand is occupied by H$_2$O 
for aqueous Mb but is coordinated with His 64 when dry.  In the latter case the Fe atom 
is in the plane of porphyrin ring in a low spin ($S = 1/2$) configuration (metMb)~\cite{feng:2001}.  The 
ligand field splits the Fe energy levels into two $e_g$ and three $t_{2g}$ orbitals, 
as illustrated in figure~\ref{fig:Heme}(b), with 
an energy gap $\Delta\approx 1.0\textrm{ eV}$ for metMb~\cite{rosca:2000}.  These orbitals split further by 0.25~eV to 
0.37~meV due to molecular orbital interactions~\cite{feng:2001}, which is difficult to detect even at 
a cryogenic temperature of $T \sim 5$~K.  
Excited states are also difficult to probe because of low $V_B$  ($< 300$~mV) used in the 
measurements (higher voltages destroyed our samples).

\begin{figure}
\centering
\includegraphics{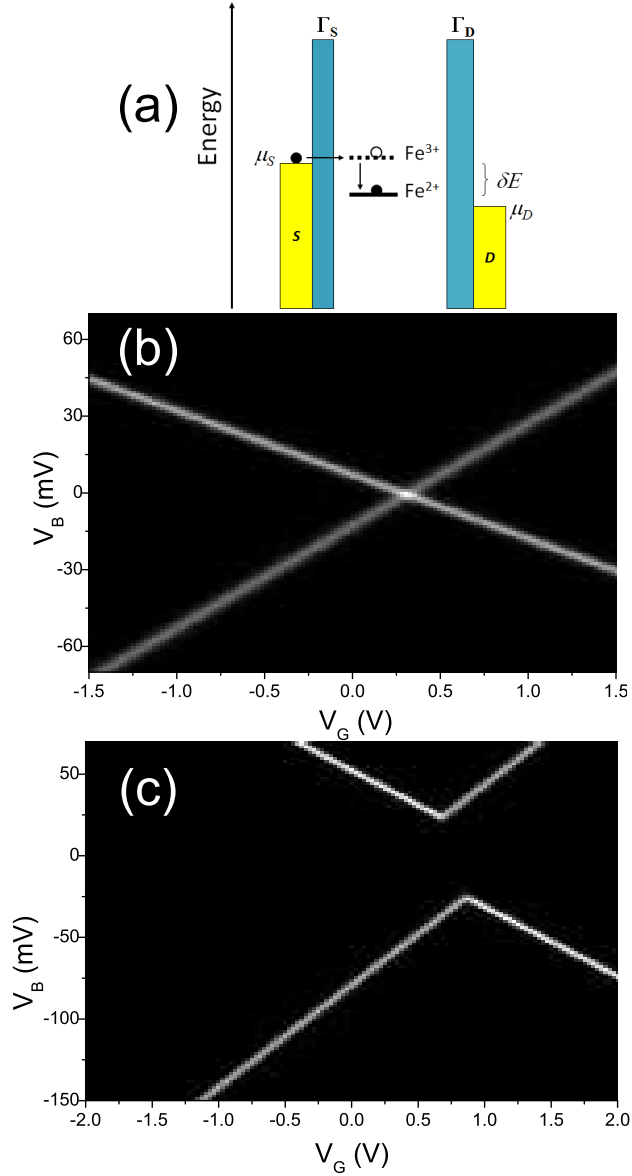}
\caption{\label{fig:Calculations} (a) Two-step electron transfer by redox center.  
$\delta E$ is the difference in conformational 
energies between the unoccupied and occupied states.  The Fermi levels of the source and drain electrodes are
indicated by $\mu_S$ and $\mu_D$, respectively. 
(b) Calculated differential conductance $dI/dV$ as a function of bias and gate voltages $V_B$ and $V_G$
 for $T = 6$~K, $\delta E = 0$~mV, $V_C = 0.307$~V, $C_D:C_S:C_G = 24:40:1$, 
tunneling rate 10~GHz (between the
 protein and the source or drain electrodes). $C_D$, $C_S$ and $C_G$ are 
the capacitances of the drain, source, and gate electrical contacts, 
respectively.  There is no two-step process in this case ($\delta E = 0$).  
(c) Calculated $dI/dV$ with a two-step process for $\delta E = 10$~mV, other 
parameters same as in (b).  See \textit{Appendix A} for description of calculation.}
\end{figure}

A Mb SET can be modeled as a single energy level device with inelastic tunneling 
processes possible due to the localized vibrational modes of the heme group.  However, 
it is known that redox reactions can cause significant conformational changes of the heme 
group~\cite{tezcan:1998} and that the associated relaxation time could range from picoseconds to 
seconds~\cite{cherepanov:2001}.  If the relaxation time is long, the electron resonantly tunnels while the 
protein remains in the intermediate state and its degeneracy point is at $V_B$ = 0.  If the 
relaxation time is short (on the order of a few ps), the protein quickly relaxes from the intermediate 
state to its ground state, whose level may fall below the Fermi level of electrodes,
as illustrated in figure~\ref{fig:Calculations}(a).
  In this case, the electron is trapped in the protein, current flow is blocked, 
and the Coulomb blockade lacks a degeneracy point.  This is further illustrated by the calculations
in figures~\ref{fig:Calculations}(b) and \ref{fig:Calculations}(c) 
performed by solving tunneling rate equations~\cite{bonet:2002,park:2003}
(see \textit{Appendix A}).  
When the two-step relaxation process does not take place, normal Coulomb triangles are 
observed in the differential conductance $dI/dV$ map.  When the relaxation process occurs,
a gap opens at the intersection of the Coulomb triangles (degeneracy point).  

It is important to note that amino acids around the heme group may complicate the analysis by acting as intermediaries 
for electron hopping processes and redox centers~\cite{shih:2008}.  
These aromatic amino acids have low frequency vibrational modes, 
such as phenylalanine (124 meV), tyrosine (79.9 meV, 103 meV, and 106 meV), and tryptophan (94.2 meV, 109 meV and 
125 meV)~\cite{wei:2008}.  

A fundamental question in analyzing our results is determining whether the protein is indeed being measured.  
Regarding the results of the bare junctions with respect to the Mb junctions, we notice
that the conductivity maps in figures \ref{fig:myo1data} and \ref{fig:myo2data} do not show
multiple degeneracy point like those of the bare junctions in figure \ref{fig:blanks}(d), and the 
energy scale (gate and bias voltages) used are completely different.  This 
is an indication that the myoglobin data come from the protein and not from
metallic islands formed in the junctions.  We note that a similar observation has been made for
break junctions used to measure single OPV-5 molecules, where gold grains remaining in the 
junction create multiple Coulomb diamonds which are superimposed on the signal
from the molecule~\cite{vanderzant:2006a}.  As to whether the heme group plays a role in 
resonant tunneling processes, we note that none of the apoMb samples' conductance showed a strong dependence on
gate voltage, while only Mb samples had this dependence.  
Another indication that the data represent the conductance of the protein are the inelastic tunneling
features of the conductance plots in figures~\ref{fig:myo1data} and \ref{fig:myo2data}.  As mentioned
above, these features correspond to the known vibrational energies of the heme group.

The effects of the protein orientation can be surmised from inelastic vibrational 
spectra.  Two pathways for electron transfer between the heme iron and electrodes 
are through the porphyrin ring or the His93 group~\cite{feng:2001}.  Transfer through the 
porphyrin ring occurs as a result of porphyrin ring’s $p$-orbital hybridization with the  
$d$-orbital of iron~\cite{feng:2001} combined with a slow porphyrin-iron structural relaxation~\cite{stavrov:2004}.  
In this case, the resonant energy level does not change during tunneling, and results 
like those in figures \ref{fig:Calculations}(b) and \ref{fig:myo2data}(a) are expected.  On the 
other hand, the dip near $V_B = 0$ and $V_G < 3$~V in Fig.~\ref{fig:myo2data}(b) 
could be due to the very fast relaxation (~ps) of the 
Fe-His bond~\cite{stavrov:2004,spiro:1988}.  The qualitative difference between 
figures~\ref{fig:myo1data}(a) and \ref{fig:myo2data}(a) 
could be due to different orientations of the heme group with respect to the electrodes.  Moreover, 
because the heme group is not at the center of the protein~\cite{garrett:1999}, certain orientations 
may cause the heme to be closer to the electrodes, thus increasing the 
probability of observing Kondo resonances.

Another complication is that the large lateral size of the break junctions
could result in several proteins with different conformations being measured in parallel, or in protein
dimers consisting of two proteins measured in series if the gap is somewhat larger than a single protein.  
We assume that conductance is dominated by the narrowest portions of the break junction since tunneling
processes usually depend exponentially on the effective width of the barriers, and therefore
it is unlikely that trimers or larger protein agglomerations will play much of a role in the conductance
data.  The data shown in figure \ref{fig:myo2data}(a) agree well with what is expected from a single
protein conductance, whereas the data in figure~\ref{fig:myo2data}(b) show indications of multiple
protein conductance (either in series or in parallel), in addition to a two-step tunneling process.  
Analyzing the data with more 
sophisticated models that take into account multiple protein conductance may be helpful in 
further understanding the conductance mechanism, but this is beyond the scope of this paper.  

\section{Conclusions}
We have presented results from a series of experiments on metallic break junctions
with apomyoglobin and myoglobin.  Some myoglobin junctions displayed single-electron transistor
behavior consistent with resonant tunneling through the heme group.  In one case, 
tunneling appears to be a result of a two-step process where the protein changes conformation
during the tunneling process.  Elastic and intelastic tunneling data are also consistent
with resonant tunneling from known vibrational modes of the protein.  Our results
underscore the importance of tunneling, as opposed to hopping, as an electron transfer mechanism
 in redox proteins.  

Future challenges include determining the spin state of a single metalloprotein via a 
SET configuration for sensor applications and using other metalloproteins, such as 
cytochrome P450, that have binding properties to specific molecules. This will require a greater
 yield of successful devices and control of the 
protein orientation and conformation on the electrodes.

\ack
The authors thank S.\ Urazhdin and his group for their help with the design of the 
electronic chips and the use of cryogenic equipment.  J.\ Gu helped with the initial 
phases of the experiment and J.\ Kabulski participated in useful discussions on protein attachemnt.  
This work was supported by the U.\ S.\ National Science Foundation (grants EPS-0554328 and EPS-1003907), the U.\ S.\ 
National Institutes of Health (grant GM-086891), and the WVNano Initiative at WVU.  
Some of the work was performed using the 
WVU Shared Research Facilities.

\appendix
\section{Rate Equations for Two-step Electron Transfer}
In the following treatment we use the same approach as in references \cite{bonet:2002} and \cite{park:2003}.  
The rate equation for the transition of an electron into an energy level in a SET is 
\begin{equation}
\frac{dP_\alpha}{dt}=\sum_\beta (\Gamma_{\beta\rightarrow\alpha}P_\beta-\Gamma_{\alpha\rightarrow\beta}P_\alpha),
\end{equation}
where $P$ is the probability that the molecule is in a specific state and $\Gamma_{\beta\rightarrow\alpha}$ and $\Gamma_{\alpha\rightarrow\beta}$
are the transition rates between $\alpha$ and $\beta$ states.  The current can be calculated by electron transfer 
through either of the two tunnel barriers corresponding to the source and drain electrical contacts.  
The two currents at equilibrium are equal.  The current through the source is 
\begin{equation}
I=I_S=e\sum_\alpha\sum_\beta\Gamma_{\alpha\rightarrow\beta}^S P_\alpha,
\end{equation}
where $e$ is the charge of the electron, $\Gamma_{\alpha\rightarrow\beta}^S P_\alpha$ is the contribution of the source to $\Gamma_{\alpha\rightarrow\beta}$
multiplied by $\pm 1$ depending on the $\Gamma_{\alpha\rightarrow\beta}$ transition contribution to the current.

Figure~\ref{fig:Calculations}(a) is a diagram representing the two-step resonant tunneling process.  The protein’s ferric state (Fe$^{+3}$) is labeled 
by (0) and the ferrous state (Fe$^{+2}$) is labeled (1).    $\Gamma_S$ and $\Gamma_D$  are the tunneling rates between the 
protein and the source and drain electrodes, respectively.  The energy difference between the ferric and ferrous states is $\delta E$. 
 We define  $P_0$ and $P_1$  as the probabilities that the level in the quantum dot (protein) are unoccupied and occupied, respectively.

For  single electron transport, the total tunneling time $\tau$ can be estimated from $\tau=e/I$, so that $\tau\sim$ps for a current
of $I\sim 10^{-7}$A.  We therefore set up the rate equations for two-step electron transfer by making the following assumptions: 
\begin{enumerate}
\item When the electron tunnels into the protein, the energy level 0 relaxes to level 1 via a structural distortion in a time $\sim\tau$.
\item The process is not reversible and fluctuations between levels 1 and 0 resulting from nuclear distortions are neglected.  
This is true at low temperatures where the thermal energy is much smaller than the vibrational energy of the heme group ($k_BT \ll E_\textrm{vib}$).
\end{enumerate}
With these assumptions, the rate of change of $P_0$ and $P_1$ can be written as
\begin{equation}
\label{eq:rate}
\frac{\partial P_0}{\partial t}=-P_0\left(\Gamma_Sf_S^0 + \Gamma_Df_D^0\right) +P_1\left[ \Gamma_S\left( 1-f_S^1\right) +\Gamma_D\left( 1-f_D^1\right)\right] =-\frac{\partial P_1}{\partial t},
\end{equation}
where $f_S^{0(1)}$ and $f_D^{0(1)}$ are the Fermi distribution functions of the source and drain electrodes, respectively, for the unoccupied
 (occupied) energy levels.  These can be written as 
\begin{eqnarray}
f^0_{S,D}=\left[ 1+\textrm{exp}\frac{\left( \mu_{N+1}-\mu_{S,D}\right) }{k_BT}\right] ^{-1}\\  
f^1_{S,D}=\left[ 1+\textrm{exp}\frac{\left( \mu_{N+1}-\delta E-\mu_{S,D}\right) }{k_BT}\right] ^{-1}
\end{eqnarray}
Here $\mu_N$ is the chemical potential of the system with $N$ electrons and $\mu_{N+1}$ is the chemical potential when an additional electron is 
added to the system.  The latter can be expressed as
\begin{equation}
\mu_{N+1}=E_0-eV_\textrm{dot}=E_0-e\frac{C_GV_G+C_SV_{SD}}{C_{total}},
\end{equation}
where $V_\textrm{dot}$ is the potential of the quantum dot, $V_{SD}$ is the source-drain potential difference, $V_G$ is the gate potential, 
$C_{total}=C_G+C_S+C_D$ is the total capacitance of th all the electrodes in the system with $C_G$, $C_S$, and $C_D$ being
the capacitances of the gate, source, and drain electrodes.  The energy $E_0\equiv eV_CC_G/C_{total}$ 
is proportional to the crossing potential $V_C$, which is the potential at which current can flow at low bias due to 
matching of the empty electron energy level with the source electrode chemical potential.  

In equilibrium, $\frac{\partial P_0}{\partial t}=\frac{\partial P_1}{\partial t}=0$ and $P_0+P_1=1$.  Combining this
equilibrium condition with equation \ref{eq:current} and the expression for the current 
\begin{equation}
I=e\left( P_1\Gamma_S\left( 1-f_S\right) -P_0\Gamma_Sf_S\right)
\end{equation}
leads to
\begin{equation}
P_0=\frac{\Gamma_S\left( 1-f_S^1\right) +\Gamma_D\left( 1-f_D^1\right) }{\Gamma_S\left( 1+f_S^0-f_S^1\right) +\Gamma_D\left( 1+f_D^0-f_D^1\right) }
\end{equation}
and
\begin{equation}
\label{eq:current}
I=e\Gamma_S\left[ 1-f_S^1-P_0\left( 1+f_S^0-f_S^1\right) \right].
\end{equation}

In figures ~\ref{fig:Calculations}(b) and (c) $dI/dV$ was calculated by numerically taking the 
derivative of the expression for $I$ in equation~\ref{eq:current} with respect to $V_{SD}$.  As in the experiments, the chemical potential of the drain was 
set to  $\mu_D=0$ (i.e., the drain contact was grounded).  Note that if  $\delta E= 0$ there is no two-step process and the result of a single-level 
quantum dot is recovered.  
\section*{References}

\bibliographystyle{iopart-num}

\begin{thebibliography}{10}
\expandafter\ifx\csname url\endcsname\relax
  \def\url#1{{\tt #1}}\fi
\expandafter\ifx\csname urlprefix\endcsname\relax\def\urlprefix{URL }\fi
\providecommand{\eprint}[2][]{\url{#2}}

\bibitem{meunier:2004}
Munier B, de~Visser S~P and Shaik S 2004 {\em Chemical Reviews\/} {\bf 104}
  3947--3980 pMID: 15352783 (\textit{Preprint}
  \eprint{http://pubs.acs.org/doi/pdf/10.1021/cr020443g})
  \urlprefix\url{http://pubs.acs.org/doi/abs/10.1021/cr020443g}

\bibitem{marcus:1985}
Marcus R~A and Sutin N 1985 {\em Biochimica et Biophysica Acta (BBA) - Reviews
  on Bioenergetics\/} {\bf 811} 265 -- 322 ISSN 0304-4173
  \urlprefix\url{http://www.sciencedirect.com/science/article/pii/030441738590014X}

\bibitem{bendall:1996}
Bendall D~S 1996 {\em Protein Electron Transfer\/} (Oxford: BIOS Publishers)

\bibitem{tao:1996}
Tao N~J 1996 {\em Phys. Rev. Lett.\/} {\bf 76}(21) 4066--4069
  \urlprefix\url{http://link.aps.org/doi/10.1103/PhysRevLett.76.4066}

\bibitem{chi:2005}
Chi Q, Farver O and Ulstrup J 2005 {\em Proceedings of the National Academy of
  Sciences of the United States of America\/} {\bf 102} 16203--16208
  (\textit{Preprint}
  \eprint{http://www.pnas.org/content/102/45/16203.full.pdf+html})
  \urlprefix\url{http://www.pnas.org/content/102/45/16203.abstract}

\bibitem{alessandrini:2006}
Alessandrini A, Corni S and Facci P 2006 {\em Phys. Chem. Chem. Phys.\/} {\bf
  8}(38) 4383--4397 \urlprefix\url{http://dx.doi.org/10.1039/B607021C}

\bibitem{alessandrini:2005}
Alessandrini A, Salerno M, Frabboni S and Facci P 2005 {\em Applied Physics
  Letters\/} {\bf 86} 133902 (pages~3)
  \urlprefix\url{http://link.aip.org/link/?APL/86/133902/1}

\bibitem{ron:2010}
Ron I, Sepunaru L, Itzhakov S, Belenkova T, Friedman N, Pecht I, Sheves M and
  Cahen D 2010 {\em Journal of the American Chemical Society\/} {\bf 132}
  4131--4140 pMID: 20210314 (\textit{Preprint}
  \eprint{http://pubs.acs.org/doi/pdf/10.1021/ja907328r})
  \urlprefix\url{http://pubs.acs.org/doi/abs/10.1021/ja907328r}

\bibitem{park:1999}
Park H, Lim A~K~L, Alivisatos A~P, Park J and McEuen P~L 1999 {\em Applied
  Physics Letters\/} {\bf 75} 301--303
  \urlprefix\url{http://link.aip.org/link/?APL/75/301/1}

\bibitem{park:2002}
Park J, Pasupathy A~N, Goldsmith J~I, Chang C, Yaish Y, Petta J~R, Rinkoski M,
  Sethna J~P, Abruna H~D, McEuen P~L and Ralph D~C 2002 {\em Nature\/} {\bf
  417} 722--725 \urlprefix\url{http://dx.doi.org/10.1038/nature00791}

\bibitem{kubatkin:2003}
Kubatkin S, Danilov A, Hjort M, Cornil J, Bredas J~L, Stuhr-Hansen N, Hedegard
  P and Bjornholm T 2003 {\em Nature\/} {\bf 425} 698 -- 701
  \urlprefix\url{http://dx.doi.org/10.1038/nature02010}

\bibitem{yu:2004}
Yu L~H, Keane Z~K, Ciszek J~W, Cheng L, Stewart M~P, Tour J~M and Natelson D
  2004 {\em Physical Review Letters\/} {\bf 93} 266802
  \urlprefix\url{http://link.aps.org/doi/10.1103/PhysRevLett.93.266802}

\bibitem{vanderzant:2006a}
van~der Zant H~S~J, Osorio E~A, Poot M and O'Neill K 2006 {\em physica status
  solidi (b)\/} {\bf 243} 3408--3412 ISSN 1521-3951
  \urlprefix\url{http://dx.doi.org/10.1002/pssb.200669185}

\bibitem{vanderzant:2006}
van~der Zant H~S~J, Kervennic Y~V, Poot M, O{'}Neill K, de~Groot Z, Thijssen
  J~M, Heersche H~B, Stuhr-Hansen N, Bjornholm T, Vanmaekelbergh D, van Walree
  C~A and Jenneskens L~W 2006 {\em Faraday Discuss.\/} {\bf 131} 347--356
  \urlprefix\url{http://dx.doi.org/10.1039/B506240N}

\bibitem{garrett:1999}
Garrett R~H and Grisham C~M 1999 {\em Biochemistry\/} (Harcourt-Brace College
  Publishers) pp 480 -- 481, 114

\bibitem{houck:2005}
Houck A~A, Labaziewicz J, Chan E~K, Folk J~A and Chuang I~L 2005 {\em Nano
  Letters\/} {\bf 5} 1685--1688 (\textit{Preprint}
  \eprint{http://pubs.acs.org/doi/pdf/10.1021/nl050799i})
  \urlprefix\url{http://pubs.acs.org/doi/abs/10.1021/nl050799i}

\bibitem{chae:2006}
Chae D~H, Berry J~F, Jung S, Cotton F~A, Murillo C~A and Yao Z 2006 {\em Nano
  Letters\/} {\bf 6} 165--168 (\textit{Preprint}
  \eprint{http://pubs.acs.org/doi/pdf/10.1021/nl0519027})
  \urlprefix\url{http://pubs.acs.org/doi/abs/10.1021/nl0519027}

\bibitem{hayashi:2008}
Hayashi T and Fujisawa T 2008 {\em Nanotechnology\/} {\bf 19} 145709
  \urlprefix\url{http://stacks.iop.org/0957-4484/19/i=14/a=145709}

\bibitem{li:2009}
Li D 2009 {\em Exploring electron transfer in myoglobin-based transistors\/}
  Ph.D. thesis West Virginia University

\bibitem{esen:2005}
Esen G and Fuhrer M~S 2005 {\em Applied Physics Letters\/} {\bf 87} 263101
  (pages~3) \urlprefix\url{http://link.aip.org/link/?APL/87/263101/1}

\bibitem{taychatanapat:2007}
Taychatanapat T, Bolotin K~I, Kuemmeth F and Ralph D~C 2007 {\em Nano
  Letters\/} {\bf 7} 652--656 (\textit{Preprint}
  \eprint{http://pubs.acs.org/doi/pdf/10.1021/nl062631i})
  \urlprefix\url{http://pubs.acs.org/doi/abs/10.1021/nl062631i}

\bibitem{ward:2008}
Ward D~R, Halas N~J and Natelson D 2008 {\em Applied Physics Letters\/} {\bf
  93} 213108 (pages~3)
  \urlprefix\url{http://link.aip.org/link/?APL/93/213108/1}

\bibitem{dong:2006}
Dong J and Parviz B~A 2006 {\em Nanotechnology\/} {\bf 17} 5124
  \urlprefix\url{http://stacks.iop.org/0957-4484/17/i=20/a=014}

\bibitem{fowler:1986}
Fowler A~B, Timp G~L, Wainer J~J and Webb R~A 1986 {\em Phys. Rev. Lett.\/}
  {\bf 57}(1) 138--141
  \urlprefix\url{http://link.aps.org/doi/10.1103/PhysRevLett.57.138}

\bibitem{azbel:1984}
Azbel M~Y, Hartstein A and DiVincenzo D~P 1984 {\em Phys. Rev. Lett.\/} {\bf
  52}(18) 1641--1644
  \urlprefix\url{http://link.aps.org/doi/10.1103/PhysRevLett.52.1641}

\bibitem{kastner:2000}
Kastner M~A 2000 {\em Ann.\ Phys.\ (Leipzig)\/} {\bf 9} 885--894

\bibitem{shih:2008}
Shih C, Museth A~K, Abrahamsson M, Blanco-Rodriguez A~M, Di~Bilio A~J, Sudhamsu
  J, Crane B~R, Ronayne K~L, Towrie M, Vlcek A, Richards J~H, Winkler J~R and
  Gray H~B 2008 {\em Science\/} {\bf 320} 1760--1762 (\textit{Preprint}
  \eprint{http://www.sciencemag.org/content/320/5884/1760.full.pdf})
  \urlprefix\url{http://www.sciencemag.org/content/320/5884/1760.abstract}

\bibitem{strachan:2005}
Strachan D~R, Smith D~E, Johnston D~E, Park T~H, Therien M~J, Bonnell D~A and
  Johnson A~T 2005 {\em Applied Physics Letters\/} {\bf 86} 043109 (pages~3)
  \urlprefix\url{http://link.aip.org/link/?APL/86/043109/1}

\bibitem{rosca:2000}
Rosca F, Kumar A~T~N, Ye X, Sjodin T, Demidov A~A and Champion P~M 2000 {\em
  The Journal of Physical Chemistry A\/} {\bf 104} 4280--4290
  (\textit{Preprint} \eprint{http://pubs.acs.org/doi/pdf/10.1021/jp993617f})
  \urlprefix\url{http://pubs.acs.org/doi/abs/10.1021/jp993617f}

\bibitem{gruia:2008}
Gruia F, Kubo M, Ye X and Champion P~M 2008 {\em Biophysical Journal\/} {\bf
  94}(6) 2252--2268

\bibitem{feng:2001}
Feng M and Tachikawa H 2001 {\em Journal of the American Chemical Society\/}
  {\bf 123} 3013--3020 pMID: 11457012 (\textit{Preprint}
  \eprint{http://pubs.acs.org/doi/pdf/10.1021/ja003088p})
  \urlprefix\url{http://pubs.acs.org/doi/abs/10.1021/ja003088p}

\bibitem{wei:2008}
Wei F, Zhang D, Halas N~J and Hartgerink J~D 2008 {\em The Journal of Physical
  Chemistry B\/} {\bf 112} 9158--9164 pMID: 18610961 (\textit{Preprint}
  \eprint{http://pubs.acs.org/doi/pdf/10.1021/jp8025732})
  \urlprefix\url{http://pubs.acs.org/doi/abs/10.1021/jp8025732}

\bibitem{tezcan:1998}
Tezcan F~A, Winkler J~R and Gray H~B 1998 {\em Journal of the American Chemical
  Society\/} {\bf 120} 13383--13388 (\textit{Preprint}
  \eprint{http://pubs.acs.org/doi/pdf/10.1021/ja982536e})
  \urlprefix\url{http://pubs.acs.org/doi/abs/10.1021/ja982536e}

\bibitem{cherepanov:2001}
Cherepanov D~A, Krishtalik L~I and Mulkidjanian A~Y 2001 {\em Biophysical
  Journal\/} {\bf 80} 1033 -- 1049 ISSN 0006-3495
  \urlprefix\url{http://www.sciencedirect.com/science/article/pii/S0006349501760845}

\bibitem{bonet:2002}
Bonet E, Deshmukh M~M and Ralph D~C 2002 {\em Phys. Rev. B\/} {\bf 65}(4)
  045317 \urlprefix\url{http://link.aps.org/doi/10.1103/PhysRevB.65.045317}

\bibitem{park:2003}
Park J 2003 {\em Electron Transport in Single Molecule Transistors\/} Ph.D.
  thesis University of California -- Berkeley

\bibitem{stavrov:2004}
Stavrov S~S 2004 {\em Biopolymers\/} {\bf 74} 37--40 ISSN 1097-0282
  \urlprefix\url{http://dx.doi.org/10.1002/bip.20039}

\bibitem{spiro:1988}
Spiro T~G 1988 {\em Biological Applications of Raman Spectroscopy\/} (Wiley and
  Sons) pp 133--215

\end{thebibliography}
\providecommand{\newblock}{}

\end{document}